\documentclass[a4paper,12pt]{article}
\usepackage{amssymb,amsfonts,amsmath,mathtext,cite,enumerate,float,color}
\usepackage{amssymb,amsthm,latexsym, braket, euscript,amscd, authblk}

\usepackage[russian,english]{babel}
\usepackage[cp1251]{inputenc}

\title{On non-commutative operator graphs generated by covariant resolutions of identity}

\author[1]{G.G.Amosov\thanks {gramos@mi.ras.ru}}

\author[1]{A.S. Mokeev\thanks {alexandrmokeev@yandex.ru}}

\affil[1]{Steklov Mathematical Institute of Russian Academy of Sciences}

\begin{document}

\maketitle

\begin{abstract} 
We study non-commutative operator graphs generated by resolutions of identity covariant with
respect to unitary representations of a compact group. Our main goal is searching for orthogonal
projections which are anticliques (error-correcting codes) for such graphs.
A special attention is paid to the covariance with respect to unitary representations of the circle group.
We determine a tensor product structure in the space of representation under which the
obtained anticliques are generated by entangled vectors.

\end{abstract}


\section{Introduction}

We study operator systems \cite{Choi} also known as non-commutative operator graphs \cite{Duan}. These are subspaces $\mathcal{V}$ in the algebra of all bounded linear operators in a Hilbert space $H$ closed under operator conjugation and containing the identity operator,
$$
 I\in {\mathcal V},\ A\in {\mathcal V}\ \Rightarrow A^*\in {\mathcal V}. 
$$
The theory of non-commutative operator graphs is closely related to the theory of quantum error correcting codes \cite{knill}. In \cite{Duan} operator graphs are offered as a non-commutative analogue of  the confusability graph of a communication channel. An operator graph $\mathcal V$ is said to be satisfying the Knill-Lafflame-Viola condition \cite{knill} if there is an orthogonal projection $P_{H_0}$ on a subspace $H_0 \subset H$ such that 
$$
P_{H_0} A P_{H_0}=c_{A}P_{H_0}, \ 
$$
with constants $c_A \in \mathbb{C}$ depending on $A\in {\mathcal V}$.

The subspace $H_0,\ dimH_0\ge 2,$ is said to be a quantum error-correcting code.
The projection $P_{H_0},\ dimP_{H_0}\ge 2,$ was called a quantum anticlique in \cite{weaver} and  we will adhere to this terminology.
There is a connection between the dimension of $\mathcal V$ and the ability to find anticliques \cite {weaver, amosovmokeev}. Recently it was revealed \cite {amo} that non-commutative operator graphs can be constructed by means of covariant resolutions of identity \cite{Hol}. The construction of codes attracts the attention of researchers who offer codes that are stable with respect to errors produced by specific quantum channels (see, f.e. \cite {W} for amplitude damping quantum channel). In parallel, the general theory is developed. We point to the work \cite {K}, where the concept of code entropy is introduced. We suppose that our work can be incorporated to the general theory of non-commutative operator graphs.

In quantum information theory an essential place is given to the entanglement phenomena. Physically, entanglement could be interpreted as an inability to describe a partial state of one particle in a system of several particles separately from other particles in this system. Mathematically, with loosing some of the details
\cite {Hol2}, an entangled vector (a pure entangled state) is a unit vector $\psi \in \mathbb{C}^d \otimes \mathbb{C}^{n}$ which cannot be factorized into a tensor product  of vectors in the factors, i.e. $\psi \neq \psi_1 \otimes  \psi_2, \ \forall \psi_1 \in \mathbb{C}^{d}, \  \psi_2 \in \mathbb{C}^{n}$. Application of the entanglement helps us to get incredible results in quantum computing and information theory. The most popular example of such results is the exponential quantum speed up in some computational problems. Quantum error correcting schemes are build with a wide usage of the entanglement \cite{Shorscheme, steanescheme}. The entanglement is the cause of the superactivation effect \cite{yard, shirokov}.

The paper is organized as follows. In Section 2 we explain how non-commutative operator graphs can be constructed from the resolutions of identity covariant with respect to a projective unitary representation of some compact group in a finite dimensional Hilbert space $H$. It is shown that anticliques can be found among spectral projections of unitary operators determining the representation. Sections 3 and 4 are devoted to the case of special unitary representations of the circle group $\mathbb T$. In Subsection 3.1 we
introduce a tensor structure in the four dimensional $H$ such that the anticliques obtained are generated by entangled vectors. The same results are given in Subsection 4.1 for higher dimensions of $H$. 

\section{Graphs generated by covariant resolutions of identity}

Let $G$ be a compact group with the Haar measure $\mu$ normalized by the condition $\mu(G)=1$.  Denote $\mathfrak {B}$ the $\sigma$-algebra generated by compact subsets of $G$. The map $B\in \mathfrak {B} \to M(B)$ from $\mathfrak {B}$ to the cone of all positive operators in a Hilbert space $H$ is said to be a generalized resolution of identity if
$$
M(\emptyset)=0,\ M(G)=I
$$
and
$$
M(\cup _jB_j)=\sum \limits _jM(B_j), \ \textit{for} \  B_j \cap B_k=\emptyset ,\ j\neq k,
$$
where the sum in the last equation converges in weak operator topology.
Consider a projective unitary representation $g \rightarrow U_g$ of the group $G$ in $H$.
A resolution of identity $M$ is said to be covariant with respect to the action of $G$ if for each $g \in G$ and $B \in \mathfrak {B}$
$$
U_g M(B) U^{*}_g=M(gB).
$$
Suppose that $H$ is finite dimensional, then
any covariant resolution of identity is known to have the form \cite {Hol}
\begin{equation}\label{resol}
M(B)= \int_{B} U_g M_0 U^{*}_g d \mu(g)
\end{equation}
where $M_0$ is some positive operator. Taking into account the condition $M(G)=I$ we conclude
\begin{equation}\label{resol}
\int \limits _GU_gM_0U_g^*d\mu (g)=I.
\end{equation}

Let us define a linear map $\mathbb E$ on the algebra of all bounded operators $B(H)$ in the Hilbert space $H$ as follows
\begin{equation}\label{project}
{\mathbb E}(A)=\int \limits _{G}U_gAU_g^*d\mu (g),\ A\in B(H).
\end{equation}
Then, ${\mathbb E}$ is a projection (a conditional expectation) to the algebra of fixed elements $\mathcal S$ with respect to the action $A\to U_gAU_g^*$.

The non-commutative operator graph $\mathcal V$ is said to be generated by a covariant resolution of identity $M$ if \cite{amo}
\begin{equation}\label{cl}
{\mathcal V}=span\{M(B),\ B\in \mathfrak {B}\}.
\end{equation}
Since $H$ is a finite dimensional by hypothesis, all linear subspaces of operators have finite dimensions. Thus, they are closed in any topology. It turns out that we don't need to
take a closure in (\ref {cl}).
If the non-commutative operator graph $\mathcal{V}$ is generated by a covariant resolution of identity, then
$$
U_g \mathcal{V} U^{*}_g=\mathcal{V}, \ \forall g \in G.
$$
It implies that such a graph is invariant with respect to the action of a group $G$.  

Consider the spectral decomposition of a unitary operators $U_g$ 
$$
U_g=\sum_{j\in J_g} a_{j} (g) P^{g}_{j}.
$$

\textbf{Proposition 1.}\textit{ Suppose that given $g\in G$ there exists $j_g\in J_g$ such that $P_{j_g}^g=P$ and $dimP\ge 2$. Then, $P$ is an anticlique for $\mathcal{V}$.}

Proof.

It follows from (\ref{resol}) that
$$
I=\int \limits_G U_gM_0U_g^{*} d\mu(g)=\int_G \left (\sum_{j\in J_g} a_{j} (g) P^{g}_{j}\right ) M_0 \left (\sum_{k\in J_g} \overline{a_{k} (g)} P^{g}_{k}\right )d\mu(g)=
$$
$$
PM_0P+ \int_G \sum_{ j,k \in J_g \times J_g \setminus [j_g,j_g] } a_{j} (g) P^{g}_{j} M_0  \overline{a_{k} (g)} P^{g}_{k}d\mu(g)=PM_0P+W
$$
Since $(P_j^g)$ are spectral projections for a fixed $g$ we get $PWP=0$. It results in $PM_0P=P\Rightarrow PM(B)P=P,\ B\in \mathfrak{B}$.
$\Box$

\section{A representation of the circle group in the four-dimensional Hilbert space.}

Suppose that $G$ is the circle group ${\mathbb T}=[0,2\pi ]$ with the operation $+/2\pi $. Let us define a unitary representation of $\mathbb T$ in
the four-dimensional Hilbert space $H$. Choose orthonormal basis $\{ e_+,h_+,e_-,h_-\}$ in $H$. Consider two two-dimensional subspaces $H_+=span\{e_+, h_+\}$ and $H_- =span\{e_-, h_-\}$, correspondent projections $P_+$ and $P_-$ satisfy the condition $P_+ +  P_- = I$. Let the unitary representation of $\mathbb T$ be defined by the formula
\begin{equation}\label{represent}
U_{\varphi }=e^{i\varphi }P_++e^{-i\varphi}P_-,
\end{equation} 
$\varphi \in {\mathbb T}$.
If the matrix $A$ represented in ordered basis $\{ e_+,h_+,e_-,h_-\}$ in such form of $2\times 2$ blocks
\begin{equation}\label{bform}
A=
\begin{pmatrix}
A_{11} & A_{12}  \\ 
A_{21}& A_{22}\\ 
\end{pmatrix}
\end{equation}
then the action of the group results in following
\begin{equation}\label{baction}
U_{\varphi}A U_{\varphi}^{*}=
\begin{pmatrix}
A_{11} &e^{2i \varphi} A_{12}  \\ 
e^{-2i \varphi}A_{21}& A_{22}\\ 
\end{pmatrix}
\end{equation}

\textbf{Lemma 1.} {\it The algebra $\mathcal S$ consisting of stationary points under the action $A \rightarrow U_{\varphi}AU_{\varphi}^{*}$ has the form
$$
\mathcal{S}=\{ A: A=P_{+}A P_{+} + P_{-}A P_{-} \}.
$$
}

\textbf{Proof.}

By the definition of stationarity $U_{\varphi}A U_{\varphi}^{*}=A, \ \forall \varphi$ we obtain from (\ref{baction}) that the blocks $A_{12},A_{21}$ must be zero.
It follows that $\mathcal S$ is the algebra of block-diagonal matrices. 

$\Box$

\textbf{Remark 1.} \textit{It immediately follows from Lemma 1 that the map ${\mathbb E}(A)=P_+AP_++P_-AP_-$
is a conditional expectation to the algebra $\mathcal S$.}

Let us consider the operator space
$$
\mathcal{A}=\{ A: A=P_{-}A P_{+} + P_{+}A P_{-} \}
$$
Every matrix $M_0$ can be represented as a sum
$$
M_0=S+A, \ S \in \mathcal{S},\ A \in \mathcal{A}. 
$$

The following two simple statements are given without proofs.

\textbf{Proposition 2.} {\it For a positive operator $M_0$ consider the graph $\mathcal{V}= span\{I, U_\varphi M_0 U_{\varphi }^{*}, \varphi \in {\mathbb T} \} $, then $P_{+}$ and $P_{-}$ are anticliques for $\mathcal{V}$ iff 
\begin{equation}\label{form}
M_0=c_1 P_{+}+c_2 P_{-}+A,\ A\in \mathcal{A}.
\end{equation}
}

\textbf{Remark 2.} \textit{For operators of the form (\ref{form})  $span\{I, U_\varphi M_0 U_\varphi ^{*}, \varphi \in \mathbb {T} \}= span\{U_\varphi M_0 U_\varphi ^{*}, \varphi \in \mathbb {T} \}$ iff $c_1=c_2$.}

\textbf{Corollary 1.} {\it Suppose that a positive operator $M_0=c I+A_0,\ A_0 \ \in \mathcal{A}$ and $c\neq 0$, then $\mathcal{V}= span\{U_\varphi M_0 U_{\varphi}^{*}, \varphi \in \mathbb {T} \} $ is the graph with anticliques $P_{+}$ and $P_{-}$.}

\textbf{Proof.}

At first, we need to prove $I\in \mathcal{A}$. Taking into account (\ref{baction}) we obtain
$$
2cI =M_0+U_{\frac{\pi}{2}}M_0U_{\frac{\pi}{2}}^*\in span\{U_\varphi M_0 U_{\varphi}^{*}, \varphi \in \mathbb T \}.
$$
A positivity of $M_0$ results in $W\in \mathcal{V} \Rightarrow W^{*} \in \mathcal{V}$. 

$\Box$

\textbf{Remark 3.} {\it All graphs generated from $M_0$ having the form defined in Corollary 1 are subgraphs of
$$
\mathcal{V}=\{ h P_{+}+q P_{-}+A: h,q \in \mathbb{C}, \ A \in \mathcal{A} \}.
$$
}

\textbf{Proposition 3.} {\it If $M_0=c I+S_0,\ c\neq 0, \  S_0 \in \mathcal{A}$ then $ span\{ U_\varphi M_0 U_\varphi ^{*}, \varphi \in {\mathbb T} \}=span\{I, F_0,G_0\} $ for $F_0=P_+ S_0 P_-,\ G_0=P_-S_0P_+$.}

\textbf{Proof.}
 
Given a vector $A \in span\{ U_\varphi M_0 U_\varphi ^{*}, \varphi \in \mathbb {T} \}$ we get
\begin{equation}\label{chep}
A=\sum_{j=1}^{d} \alpha_{j} U_{\phi_j} M_0  U^{*}_{\phi_j}=c\left (\sum_{j=1}^{d} \alpha_{j}\right )I+\left (\sum_{j=1}^{d} \alpha_{j}e^{i \phi_j}\right )F_0+\left (\sum_{j=1}^{d} \alpha_{j}e^{-i \phi_j}\right )G_0.
\end{equation}
Thus, $span\{ U_\varphi M_0 U_\varphi^{*}, \varphi \in \mathbb {T} \} \subseteq span\{I, F_0,G_0\}$.

Substituting $(\alpha _1=\alpha_2=1,\varphi _1=0,\varphi _2=\pi )$, $(\alpha _1=-\alpha _2=1, 
\varphi _1=0,\varphi _2=\pi)$ and $(\alpha _1=1,\alpha _2=i,\ \varphi _1=0,\varphi _2=\frac {\pi }{2})$ to (\ref {chep}) we get $A=2cI$, $A=2(F_0+G_0)$ and $A=(1+i)cI+2G_0$ respectively. The result follows.

$\Box $

\textbf{Remark 4.} \textit{It immediately follows from Proposition 3 that
$$
\mathcal{V}=span\{ U_\varphi M_0 U_\varphi ^{*}, \varphi \in {\mathbb T} \}
$$ 
is a non-commutative operator graph iff $G_0=hF_0^*$ for some $h\in {\mathbb C}$.}

\textbf{Proposition 4.} {\it Let $M_0=Q$ be an orthogonal projection for which $\mathcal{V}= span\{U_\varphi Q U_{\varphi}^{*}, \varphi \in \mathbb {T} \} $ is a graph and it has the anticliques $P_{+}$ and $P_{-}$. Then, in the ordered basis $\{ e_+,h_+,e_-,h_-\}$ either $Q=I$ or
$$
Q=
\begin{pmatrix}
\frac{1}{2} & 0 &\tau e^{iz_1} &\sqrt{\frac{1}{4} -\tau ^2}e^{iz_2} \\ 
0& \frac{1}{2}  & \sqrt{\frac{1}{4} -\tau ^2}e^{iz_3} &\tau e^{iz_4} \\ 
\tau e^{-iz_1} &\sqrt{\frac{1}{4} -\tau ^2}e^{-iz_3} &\frac{1}{2}  & 0\\ 
 \sqrt{\frac{1}{4} -\tau ^2}e^{-iz_2} &\tau e^{-iz_4} & 0 & \frac{1}{2} 
\end{pmatrix}
$$
$z_1,z_2,z_3,z_4 \in \mathbb{R}, 0\le \tau \le \frac{1}{2},$
$$
z_3-z_1=z_4-z_2+\pi +2\pi k.
$$
}

{\bf Remark 5.} {\it If $Q$ satisfies the conditions of Proposition 4, then the same holds true for $I-Q$.}

{\bf Remark 6.} {\it Let $Q\neq I$, put 
$$
\xi _Q=\left (\frac {1}{2},0,\tau e^{-iz_1},\sqrt {\frac {1}{4}-\tau ^2}e^{-iz_2}\right )^T,
$$
$$
\eta _Q=\left (0,\frac {1}{2},\tau e^{-iz_1},\sqrt {\frac {1}{4}-\tau ^2}e^{-iz_2}\right )^T,
$$ 
$$
\xi _{I-Q}=\left (\frac {1}{2},0,\sqrt {\frac {1}{4}-\tau ^2}e^{-iz_3},\tau e^{-iz_4}\right )^T,
$$
$$
\eta _{I-Q}=\left (0,\frac {1}{2},\sqrt {\frac {1}{4}-\tau ^2}e^{-iz_3},\tau e^{-iz_4}\right )^T.
$$ 
Then,
$$
H_Q=QH=\{\lambda \xi _Q+\mu \eta _Q,\ \lambda ,\mu \in \mathbb {C}\},
$$
$$
H_{I-Q}=(I-Q)H=\{\lambda \xi _{I-Q}+\mu \eta _{I-Q},\ \lambda ,\mu \in \mathbb {C}\},
$$

}
\textbf{Proof.}

It follows from Corollary 1 that $Q$ should have the following form
$$
Q=
\begin{pmatrix}
c & 0 &a &d \\ 
0& c & q &b \\ 
\bar{a} & \bar{q} & c & 0\\ 
\bar{d} & \bar{b} & 0 & c
\end{pmatrix}.
$$
Since $Q=Q^{2}$ we get
$$
\begin{pmatrix}
c & 0 &a &d \\ 
0 & c & q &b \\ 
\bar{a} & \bar{q} & c & 0\\ 
\bar{d} & \bar{b} & 0 & c
\end{pmatrix}
=
\begin{pmatrix}
c^2+|a|^2+|d|^2 & a\overline q+d\overline b &2ca &2cd \\ 
\overline a q+\overline d b & c^2+|b|^2+|q|^2 & 2cq &2cb \\ 
2c\bar{a} & 2c\bar{q} & c^2+|q|^2+|a|^2 & \overline a d+\overline q b\\ 
2c\bar{d} & 2c\bar{b} & a\overline d+q\overline b & c^2+|b|^2+|d|^2
\end{pmatrix}
$$
If some of the entries $a,b,d,q$ are not equal to $0$, then $c=\frac{1}{2} $. By this way,
$$|a|^2=|b|^2=c-c^2-|d|^2=\frac{1}{4}-|d|^2$$ 
$$|d|^2=|q|^2=c-c^2-|a|^2=\frac{1}{4}-|a|^2.$$
Denote $\tau=|a|$ and let $z_1,z_2,z_3,z_4$ be the arguments of $a,d,q,b$. Equations
$$
 \bar{a}q+\bar{d}b=0
$$
$$
a\bar{d}+q\bar {b}=0
$$
take the form
$$
\tau \sqrt{\frac{1}{4} -\tau^2}(e^{i(z_3-z_1)}+e^{i(z_4-z_2)})=0
$$
$$
\tau \sqrt{\frac{1}{4} -\tau^2}(e^{i(z_1-z_2)}+e^{i(z_3-z_4)})=0
$$
It holds without constraints on the arguments if $\tau=0$ or $\tau=\frac{1}{2}$. Otherwise
$$
z_3-z_1=z_4-z_2+(2k+1)\pi
$$ 
$$
z_1-z_2=z_3-z_4+(2l+1)\pi
$$
with arbitrary parameters $k,l \in \mathbb{Z}$. Note that the two last claims are
equivalent.

$\Box$

\subsection {A tensor product structure}

It follows from Proposition 4 that if an orthogonal projections $Q\neq I$ it generates the graph only under
the condition $dimQ=2$.  Let us split $H=\mathbb{C}^{4}$ into two tensor factors $H=\mathbb{C}^{2} \otimes \mathbb{C}^{2}$ in such way that the subspace $H_Q=QH$ contains only separable
vectors
\begin{equation}\label{separ}
H_Q=x\otimes {\mathbb C}^2.
\end{equation}
Then, there exists $y\in {\mathbb C}^2,\ (x,y)=0,$ such that
\begin{equation}\label{separ2}
H_{I-Q}=(I-Q)H=y\otimes {\mathbb C}^2
\end{equation}

{\bf Proposition 5.} {\it If $\tau \neq 0$ or $\frac {1}{2}$ all possible bases of $H_{\pm}=P_{\pm}H$ consist of entangled vectors.
Moreover, if $\tau =\frac {1}{2\sqrt 2}$ they are maximally entangled.}

{\bf Proof.}

It follows from Remark 7 that
$$
\xi _Q=\frac {1}{2}e_++\tau e^{-iz_1}e_-+\sqrt {\frac {1}{4}-\tau ^2}e^{-iz_2}h_-,
$$
$$
\eta _Q=\frac {1}{2}h_++\tau e^{-iz_1}e_-+\sqrt {\frac {1}{4}-\tau ^2}e^{-iz_2}h_-,
$$
$$
\xi _{I-Q}=\frac {1}{2}e_++\sqrt {\frac {1}{4}-\tau ^2}e^{-iz_3}e_-+\tau e^{-iz_4}h_-
$$
and
$$
\eta _{I-Q}=\frac {1}{2}h_++\sqrt {\frac {1}{4}-\tau ^2}e^{-iz_3}e_-+\tau e^{-iz_4}h_-.
$$
Identifying the elements by the rule $\xi _Q=x\otimes x,\ \eta _Q=x\otimes y,\ \xi _{I-Q}=y\otimes y,\ 
\eta _{I-Q}=y\otimes x,$ we get
$$
e_+=\frac {2}{\sqrt {\frac {1}{4}-\tau ^2}e^{iz_3}+\tau e^{iz_4}}\left (\sqrt {\frac {1}{4}-\tau ^2}e^{iz_3}x\otimes x+\tau e^{iz_4}y\otimes y\right ),
$$
$$
h_+=\frac {2}{\sqrt {\frac {1}{4}-\tau ^2}e^{iz_3}+\tau e^{iz_4}}\left (\sqrt {\frac {1}{4}-\tau ^2}e^{iz_3}x\otimes y+\tau e^{iz_4}y\otimes x\right ).
$$

$\Box $

\section {A generalization for higher dimensions.}

In this section the dimension $d$ supposed to be at least $2$. Let $P_s, 1 \le s \le d,$ be orthogonal projections of the dimension $\dim P_s = d$ acting on $\mathbb{C}^{d} \otimes \mathbb{C}^{d}$ and $\sum_{s=1}^{d}P_s=I_{d} \otimes I_{d}$. A multidimensional analogue of representation (\ref {represent}) for the circle group $\mathbb T$ is the following
\begin{equation}\label{re}
\varphi \rightarrow U_{\varphi}= \sum_{s=1}^{d}e^{i \varphi s}P_s.
\end{equation}
Consider two operator spaces
$$
\mathcal{S}_{d}=\{ A: A= \sum_s P_s A P_s \}
$$
and
$$
\mathcal{A}_{d}=\{ A: A=\sum_{ s\neq k} P_k A P_s \}.
$$
Following the same way as in Lemma 1 we see that  the algebra of stationary points with respect to the action $A\to U_{\varphi }AU_{\varphi}^*$ is $\mathcal{S}_{d}$. Moreover, the following theorem holds true.

\textbf{Theorem.} {\it  Given a positive operator $M_0=cI+A_0, \ c>0,\ A_0 \in \mathcal{A}_{d}$ the operator space 
${\mathcal V}=span\{U_\varphi M_0U_\varphi ^*,\ \varphi \in {\mathbb T}\}$ is a non-commutative operator graph.
Moreover, the projections $\{P_s,\ 1\le s\le d\}$ are anticliques for $\mathcal V $.

}

\textbf{Proof.}

Consider the conditional expectation ${\mathbb E}$ to the algebra of stationary points $\mathcal S$
defined by the formula
\begin{equation}\label{id}
{\mathbb E}(A)=\sum \limits _{s=1}^dP_sAP_s.
\end{equation}
The same projection can be represented as (\ref {project})
\begin{equation}\label{id2}
{\mathbb E}(A)=\frac {1}{2\pi}\int \limits _0^{2\pi}U_{\varphi}AU_{\varphi }^*d\varphi
\end{equation}
Substituting $A=M_0$ to (\ref {id2}) we get
$$
{\mathbb E}(M_0)=cI.
$$
It results in $I\in \mathcal V$. On the other hand, $A\in {\mathcal V}\Rightarrow A^*\in {\mathcal V}$
by a construction of $\mathcal V$. Now the result follows from Proposition 1.

$\Box$

\subsection {A tensor product structure}

Suppose that $H={\mathbb C}^d\otimes {\mathbb C}^d$. Let $\ket{k}, \ 1 \le k \le d$ denote elements of some orthonormal basis in $\mathbb{C}^{d}$. Consider the generalized Bell states \cite{Bennett} in $H$ defined by
\begin{equation}\label{key}
\ket{\psi_ {sn}}=\frac{1}{\sqrt{d}} \sum_{k=1}^{d} e^{\frac{2\pi i s k}{d}}\ket{k}\ket{k-n\  mod\  d},
\end{equation}
$1\le s,n\le d$. Let $P_s, \ 1 \le s \le d$ be the projections on the subspaces
$$
H_s=span \{ \ket{\psi_{sn}}, \ 1 \le n \le d \}.
$$

In order to find an analogue of the projection determined in Proposition 4 consider a set of projections in $H=\mathbb{C}^{d}\otimes\mathbb{C}^{d}$ of the 
form
$$
Q_j=\sum \limits _{k=1}^d\ket {j}\ket {j-k\ mod\ d}\bra {j-k\ mod\ d}\bra {j},
$$
$1\le j\le d$.

\textbf{Corollary 2.} {\it Given $j,\ 1\le j\le d,$ the projection $Q_j$ generates the graph $\mathcal{V}_j= span\{ U_\varphi Q_j U_\varphi^{*}, \varphi \in {\mathbb T} \} $ for which the projections $\{P_{s},\ 1 \le s \le d\}$ are anticliques.}

\textbf{Proof.}

The subspace $H_{Q_j}=Q_jH$ is a linear envelope of unit vectors
$$
\eta _k^j=\ket {j}\ket {j-k\ mod\ d},\ 1\le k\le d.
$$
Hence,
$$
\braket {\psi _{sn} | \eta _k^j}=0
$$
for $k\neq n$ and
$$
\braket {\psi _{sn} | \eta _n^j }=\frac {1}{\sqrt d}e^{-\frac{2\pi i s j}{d}},
$$
$1\le s,j\le d$.
Thus,
$$
P_s\ket {\eta _k^j}=\frac {1}{\sqrt d}e^{-\frac {2\pi is j}{d}}\ket {\psi _{sk}}.  
$$
Applying (13) to $Q_j$ and taking into account that
$$
\sum \limits _{s,k=1}^d\ket {\psi _{sk}}\bra {\psi _{sk}}=I
$$
we obtain
$$
{\mathbb E}(Q_j)=\frac {1}{d}I.
$$
Now the result follows from Theorem.

$\Box $

\section{Conclusion}

We consider non-commutative operator graphs generated by resolutions of identity covariant with
respect to finite-dimensional projective unitary representations of compact groups. The principal example is given by
different unitary representations of the circle group. The representations resulting in 
the entanglement of separable vectors are constructed. In the case, it is shown that the spectral projections of unitary operators generating the representation become anticliques (error-correcting codes) for the graph. We plan to extend our construction to the non-commutative case. The first candidate should be the discrete Heisenberg-Weyl group for which the codes were constructed in \cite {amosovmokeev}.

\section*{Acknowledgments} The authors are extremely grateful to the anonymous referee for a careful reading of the text and many fruitful remarks. This work is supported by the Russian Science Foundation under grant 17-11-01388 and
performed in Steklov Mathematical Institute of Russian Academy of Sciences.

\end{document}